\documentclass[aps,prd,onecolumn,superscriptaddress,floats,amsfonts,amsmath,amssymb,linenumbers,12pt]{revtex4}

\pdfoutput=1
\usepackage{graphicx}
\usepackage{epstopdf}
\usepackage[active]{srcltx}
\usepackage{bm}
\usepackage[]{latexsym}

\usepackage{color}

\usepackage{caption2}

\usepackage{float}

\usepackage{lineno}

\begin{document}

\restylefloat{figure}

\newcommand*{\st}[1]{\textbf{\color{red}{*} #1 * }}

\title{On Possibility of Determining Neutrino Mass Hierarchy by the Charged-Current and Neutral-Current Events of Supernova Neutrinos in Scintillation Detectors}


\author{Fei-Fan Lee}
\affiliation{Department of Physics, Jimei University, 361021, Xiamen, Fujian province, P. R. China}

\author{Feng-Shiuh Lee}
\affiliation{Department of Electrophysics, National Chiao Tung University, Hsinchu, 300, Taiwan}

\author{Kwang-Chang Lai}
\email{kcl@mail.cgu.edu.tw}
\affiliation{Center for General Education, Chang Gung University, Kwei-Shan, Taoyuan, 333, Taiwan}




\begin{abstract}

One of the unresolved mysteries in neutrino physics is the neutrino mass hierarchy. We present a new method to determine neutrino mass hierarchy by comparing the events of inverse beta decays (IBD), $\bar{\nu}_e + p\rightarrow n + e^+$, and neutral current (NC) interactions, $\nu(\overline{\nu}) + p\rightarrow\nu(\overline{\nu}) + p$, of supernova neutrinos from accretion and cooling phases in scintillation detectors. Supernova neutrino flavor conversions depend on the neutrino mass hierarchy. On account of Mikheyev-Smirnov-Wolfenstein effects, the full swap of $\bar{\nu}_e$ flux with the $\bar{\nu}_x$ ($x=\mu,~\tau$) one occurs in the inverted hierarchy, while such a swap does not occur in the normal hierarchy. In consequence, the ratio of high energy IBD events to NC events for the inverted hierarchy is higher than in the normal hierarchy. Since the luminosity of $\bar{\nu}_e$ is larger than that of $\nu_x$ in accretion phase while the luminosity of $\bar{\nu}_e$ becomes smaller than that of $\nu_x$ in cooling phase, we calculate this ratio for both accretion and cooling phases. By analyzing the change of this event ratio from accretion phase to cooling phase, one can determine the neutrino mass hierarchy.

\vspace{3mm}


\end{abstract}

\maketitle

\section{Introduction}

Supernovae (SNe) are among the most powerful sources of neutrinos in our Universe. During a supernova explosion, $99\%$ of the emitted energy ($\sim10^{53}$  erg) is released by neutrinos and antineutrinos of all favors, with energy ranging from several to a few tens MeV, which play the role of  astrophysical messengers, escaping almost unimpeded from the supernova core. The supernova neutrino flux has been extensively studied as a probe of both fundamental neutrino properties and core-collapse physics. Therefore,  Observing the supernova neutrino signal would enable a wide range of opportunities, both in astrophysics and in particle physics.

Based on various oscillation experiments with atmospheric, solar, and terrestrial neutrinos \cite{GonzalezGarcia:2007ib}, a considerable progress has been achieved in constraining the neutrino mixing parameters \cite{Maki,Pontecorvo}. The flavor states $\nu_e$, $\nu_\mu$, and $\nu_\tau$ are now well recognized to be superpositions of the vacuum mass eigenstates $\nu_1$, $\nu_2$, and $\nu_3$ \cite{Strumia:2006db}. Thanks to many successful experiments, the three neutrino flavor mixing angles, $\theta_{12}$, $\theta_{23}$, and $\theta_{13}$,  and two mass-squared differences, $\Delta^2_{21}=m^2_2-m^2_1$ and $\Delta^2_{31}=m^2_3-m^2_1$ are well constrained, whereas the sign of $\Delta^2_{31}$, i.e.,  the neutrino mass hierarchy, is still unknown. To determine the neutrino mass hierarchy, recent efforts include works based on reactor neutrinos \cite{Petcov:2001sy,Ge:2012wj,Li:2013zyd,Capozzi:2013psa} different baseline experiments \cite{Ishitsuka:2005qi}, Earth matter effects on supernova neutrino signal \cite{Lunardini:2003eh,Dasgupta:2008my}, spectral swap of SN neutrino flavors \cite{Duan:2007bt}, rise time of SN $\nu_e$ light curve \cite{Serpico:2011ir}, $\nu_e$ and $\bar{\nu}_e$ light curves on the early accretion phase \cite{Chiu:2013dya}, analysis of meteoritic SN material \cite{Mathews:2011jq}, and detection of atmospheric neutrinos in sea water or ice \cite{Winter:2013ema}. Among them, those works using supernova neutrinos are particularly interesting because of the interplay between intrinsic properties of massive neutrinos and the mechanism of SN explosions.

Historically, the detection of neutrinos from SN1987A \cite{Hirata:1987, Bionta:1987} has motivated a huge amount of theoretical works in both SN physics and neutrino physics. Therefore, many proposals to identify neutrino mass hierarchy by studying neutrinos from galactic SNe have been proposed. Originating from deep inside the SN core, neutrinos can experience significant flavor transitions on their way to the terrestrial detectors. Neutrino flavor conversions arising from the Mikheyev-Smirnov-Wolfenstein (MSW) effect \cite{Wolfenstein:1978,Mikheev:1985} are sensitive to neutrino mass hierarchy. Additionally, it has been pointed out that the collective neutrino oscillation \cite{Dasgupta:2009mg,Dasgupta:2010ae,EstebanPretel:2007ec,Raffelt:2007cb,Hannestad:2006nj,Dasgupta:2007ws,Choubey:2010up,Duan:2005cp,Mirizzi:2010uz} (see \cite{Duan:2010bg} for a review) results from the coherent $\nu-\nu$ forward scatterings in the deep region of the core where neutrino densities are large and may lead to collective pair flavor conversion $\nu_e\bar{\nu}_e\leftrightarrow\nu_x\bar{\nu}_x$ ($x=\mu,~\tau$) over the entire energy range. However, unlike the status of MSW effects, consensus on collective flavor transitions has not yet been reached. To avoid digression to diverse scenarios of the collective effect, we assume that MSW effect dominates the flavor conversions when SN neutrinos propagate outwards.

The interactions of SN neutrinos with atomic nuclei and free protons are utilized to resolve the neutrino mass hierarchy in most of the methods. The inverse beta decay (IBD), $\bar{\nu}_e+p\rightarrow n+e^+$, is the major interaction channel for neutrino detection in the water Cherenkov and liquid scintillation detectors. On the other hand, the liquid argon detector has a good sensitivity to $\nu_e$ via charged-current interactions. Because the threshold of visible energy in a liquid scintillation detector can be as low as $0.2$ MeV \cite{Alimonti:2002} by controlling the abundance of $^{14}{\rm C}$, the neutral-current (NC) interactions, $\nu+p\rightarrow\nu+p$, will give rise to a large number of events in a channel other than IBD in this case and become very important. As a result, the detection of other species of SN neutrinos was proposed by measuring NC interactions \cite{Beacom:2002hs,Dasgupta:2011wg}.

Inspired by the capability of detecting thousands of neutrino events from a galactic supernova with next-generation scintillation detectors, we proposed to identify the neutrino mass hierarchy by comparing IBD and NC interactions inside the scintillators \cite{Lai:2016yvu}. In the previous work, a set of specific mean energies of different flavors and luminosity equipartition between flavors for SN neutrinos are adopted. Instead of a unique scenario, models with different sets of mean energies and partitions of luminosities are explored in this work and the evolution with time of SN neutrinos is also accounted for in calculating IBD and NC events in the scintillation detector. We study how the way that the IBD and NC events change with time during a SN explosion is related to the neutrino mass hierarchy.

The paper is organized as follows. In Sec. II, we briefly review the flavor transitions of SN neutrinos as they propagate outward from deep inside a SN and traverse the Earth medium to reach the detector and describe the supernova neutrino fluence in our calculation. In Sec. III, we describe the event calculation for inverse beta decay and neutral current inside liquid scintillation detectors and briefly discuss interactions of SN neutrinos in the detectors. Then, in Sec. IV, we define a ratio R of total IBD events to total NC events and present our calculations for considered parameter space with statistical uncertainties addressed. Finally, in Sec. V we summarize our results and conclude.

\section{Supernova Neutrino Fluence}

\subsection{Primary Neutrino Fluence}

A SN neutrino burst lasts for $\Delta t\approx10{\rm s}$ and includes all six flavors of neutrinos. The total gravitational binding energy released in the explosion is ${\mathcal E}\approx10^{53}~{\rm erg}$. The neutrino flavors $\nu_\mu$, $\nu_\tau$ and their antiparticles have similar interactions and thus similar average energies and fluences. Thus the total energy is divided as ${\mathcal E}={\mathcal E_{\nu_e}}+{\mathcal E_{\bar{\nu}_e}}+4{\mathcal E_{\nu_x}}$. In this work, the condition of equipartition of energies and luminosities among the primary neutrino flavors, ${\mathcal E_{\nu_e}}\approx{\mathcal E_{\bar{\nu}_e}}\approx{\mathcal E_{\nu_x}}$ and ${\mathcal L_{\nu_e}}\approx{\mathcal L_{\bar{\nu}_e}}\approx{\mathcal L_{\nu_x}}$, is relaxed. The primary SN neutrino energy spectrum is typically not purely thermal. We adopt the Keil parametrization \cite{Keil:2002in} for the neutrino fluence
\begin{equation}
F^0_\alpha(E)  =  \frac{\Phi_\alpha}{<E_\alpha>}\frac{(1+\eta_\alpha)^{(1+\eta_\alpha)}}{\Gamma(1+\eta_\alpha)}\left(\frac{E}{<E_\alpha>}\right)^{\eta_\alpha} 
                        \exp\left[-(\eta_\alpha+1)\frac{E}{<E_\alpha>}\right],
\end{equation}
where $\Phi_\alpha={\mathcal E}_{\alpha}/<E_\alpha>$ is the time-integrated flux, $<E_\alpha>$ is the average neutrino energy, and $\eta_\alpha$ denotes the pinching of the spectrum. In our calculation, we take $\eta_\alpha=3$ for all flavors.
If flavor conversions do not occur during the propagations of neutrinos from the SN core to the Earth, a SN at a distance $d$  thus yields a neutrino fluence
\begin{equation}
F_\alpha = \frac{F^0_\alpha}{4\pi d^2} 
               = \frac{2.35\times 10^{13}}{\rm cm^2 MeV}\frac{{\mathcal E}_\alpha}{d^2}\frac{E^3}{<E_\alpha>^5}\exp\left(-\frac{4E}{<E_\alpha>}\right), \label{fluence}
\end{equation}
with ${\mathcal E}_\alpha$ in units of $10^{52}$ ${\rm erg}$, $d$ in 10 ${\rm kpc}$, and energies in ${\rm MeV}$.
For the numerical evaluations, we take a representative supernova at the Galactic center region with $d=10 ~{\rm kpc}$, and a total energy output of ${\mathcal E}=3\times10^{53}~{\rm erg}$. Further, different sets of the average energies, $(<E_{\nu_e}>, <E_{\bar{\nu}_e}>, <E_{\nu_x}>)$, will be taken in our calculation. In addition to $<E_{\nu_e}>=12~\rm MeV$, $<E_{\bar{\nu}_e}>=15~\rm MeV$, $<E_{\nu_x}>=18~\rm MeV$, more hierarchical values, $<E_{\nu_e}>=10~\rm MeV$, $<E_{\bar{\nu}_e}>=15~\rm MeV$, $<E_{\nu_x}>=24~\rm MeV$, and more degenerate values, $<E_{\nu_e}>=12~\rm MeV$, $<E_{\bar{\nu}_e}>=14~\rm MeV$, $<E_{\nu_x}>=16~\rm MeV$, will also be considered.
\label{fluence}

\subsection{Neutrino Fluence on Earth}

As neutrinos propagate outwards from deep inside a SN and finally reaches the Earth, their flavor contents are modified by the MSW effect. The fluxes of $\nu_e$ and $\bar{\nu}_e$ arriving at the detector can be written as:
\begin{eqnarray} 
F_e            & = &  F^0_x,   \label{eNH}  \\
F_{\bar{e}} & = &  (1-\bar{P}_{2e}) F^0_{\bar{e}} + \bar{P}_{2e} F^0_{\bar{x}}, \label{ebarNH}
\end{eqnarray}
for the normal hierarchy, and
\begin{eqnarray} 
F_e            & = & P_{2e} F^0_e + (1-P_{2e}) F^0_x,  \label{eIH} \\
F_{\bar{e}} & = & F^0_{\bar{x}}, \label{ebarIH}
\end{eqnarray}
for the inverted hierarchy \cite{Dighe:1999bi}. Here $P_{2e}$ ($\bar{P}_{2e}$) is the probability that a mass eigenstate $\nu_2$ ($\bar{\nu}_2$) is observed as a $\nu_e$ ($\bar{\nu}_e$) since neutrinos arrive at the Earth as mass eigenstates. We do not consider the  regeneration factor due to the Earth matter effect and thus take  $P_{2e}=\sin^2\theta_{12}$ in this work. From Eqs. (\ref{eNH}) to (\ref{ebarIH}), it is shown that, in the normal hierarchy, $\nu_e$ completely comes from $\nu_x^{0}$ from the source while $\bar{\nu}_e$ comes from both $\bar{\nu}_e^{0}$ and $\bar{\nu}_x^{0}$. On the other hand, in the inverted hierarchy, $\nu_e$ comes from both $\nu_e^{0}$ and $\nu_x^{0}$ while $\bar{\nu}_e$ completely comes from $\bar{\nu}_x^{0}$.

For the rest of flavors, the condition of flux conservation gives
\begin{eqnarray}
4F_x & = & F^0_e + F^0_{\bar{e}} + 4F^0_x - F_e - F_{\bar{e}}  \nonumber \\
         & = & F^0_e + \bar{P}_{2e} F^0_{\bar{e}} + (3-\bar{P}_{2e}) F^0_x,
\end{eqnarray}
and
\begin{eqnarray}
4F_x & = & F^0_e + F^0_{\bar{e}} + 4F^0_x - F_e - F_{\bar{e}} \nonumber \\
         & = & (1-P_{2e}) F^0_e + F^0_{\bar{e}} + (2+P_{2e}) F^0_x,
\end{eqnarray}
for the normal and inverted hierarchies, respectively.


\section{Events of Inverse Beta Decay and Neutral Current Interaction inside Scintillation Detectors}

In scintillation detectors, inverse beta decays (IBD) are the most dominant interactions. IBD events are obtained in scintillation detectors by measuring the positron energy deposit. The observed event spectrum and total number of IBD events are given by
\begin{eqnarray}
&& \left(\frac{dN}{dE_{e^+}}\right)=N_p\cdot\int dE_\nu\frac{dF_{\bar{e}}}{dE_\nu}\cdot\frac{d\sigma_{\rm IBD}(E_\nu,~E_{e^+})}{dE_{e^+}}, \label{IBDspec} \\
&& N_{{\rm IBD}}=N_{e^+}=N_p\cdot\int_{E_{\rm min}}^\infty dE_\nu\frac{dF_{\bar{e}}}{dE_\nu}\cdot\sigma_{\rm IBD}(E_\nu),
\end{eqnarray}
where $N_p$ is the number of the target protons in the detector and cross section $\sigma_{\rm IBD}(E_\nu)$ is taken from \cite{Strumia:2003zx}. The minimum neutrino energy for generating IBD interaction is $E_{\rm min}=1.8~{\rm MeV}$.


Inside the scintillation detector, the yield of $\nu p$ elastic scatterings is also comparable to that of IBD due to the large number of free protons \cite{Dasgupta:2011wg}. The observed event spectrum is given as
\begin{equation}
\frac{dN}{dT'}=\frac{N_p}{dT'/dT}\int_{E_{\nu,{\rm min}}}^\infty dE_\nu\frac{dF_{\rm tot}}{dE_\nu}\frac{d\sigma_{\nu p}(E_\nu, T)}{dT},
\end{equation}
where $F_{\rm tot} \equiv F_e+F_{\bar{e}}+4F_x$  is the total fluence of the SN neutrinos and $T$ is the recoil kinetic energy of protons which are scattered by SN neutrinos. To produce a proton recoil energy $T$ requires a minimum neutrino energy $E_{\nu,{\rm min}}=\sqrt{m_p T/2}$, with $m_p$ the proton mass. In other words, a neutrino of energy $E_\nu$ can produce a proton recoil energy between $0$ and $T_{\rm max}=2E^2_\nu/m_p$. These protons are slow hence they are detected with quenched energies $T'<T$. The proton recoil energy $T$ is mapped to an electron-equivalent quenched energy $T^{\prime}$ through the quenching function
\begin{equation}
T^{\prime}(T)=\int_0^T\frac{dT}{1+k_B<dT/dx>},
\end{equation}
where $k_B$ is Birks constant \cite{Birks:1951}. The number of NC events is then given by 
\begin{eqnarray}
N_{\rm NC} & = & N_p\cdot\int_{T_{\rm min}}^\infty\int_{E_{\nu,{\rm min}}}^\infty \frac{dF_{\rm tot}}{dE_\nu}\cdot\frac{d\sigma_{\nu p}(E_\nu, T)}{dT}dE_\nu dT, 
\end{eqnarray}
where the differential cross section, $d\sigma_{\nu p}/dT$, is taken from \cite{Beacom:2002hs,Dasgupta:2011wg}.We point out that not all signals within the energy range of proton recoils are taken into account. Since the scintillator is made of hydrocarbon, a natural isotope of the carbon, $^{14}{\rm C}$, decays into $^{14}{\rm N}$, emitting electrons below $0.2$ MeV with a high rate. Below this energy, the signal is flooded by very low energy electrons. Therefore, a threshold of $T^{\prime}_{\rm min}=0.2~{\rm MeV}$ is set for recording the signal. The threshold of $T^{\prime}_{\rm min}=0.2~{\rm MeV}$ is converted to the threshold of proton recoil energy $T_{\rm min}$, e.g. $T_{\rm min}=0.93~{\rm MeV}$ for JUNO detector.

Besides IBD and NC signals, the interactions between SN neutrinos and scintillation materials also happen in other various reaction channels: (1) the elastic neutrino-electron scattering $\nu + e^- \rightarrow\nu + e^-$, (2) the charged-current $\nu_e$ interaction $\nu_e + {^{12}{\rm C}}\rightarrow {^{12}{\rm N}}_{\rm g.s.} + e^-$, (3) the charged-current $\bar{\nu}_e$ interaction $\bar{\nu}_e + {^{12}{\rm C}}\rightarrow {^{12}{\rm B}}_{\rm g.s.} + e^+$, (4) proton knockouts \cite{Lujan-Peschard:2014lta} $\nu (\bar{\nu})+ {^{12}{\rm C}}  \rightarrow  {^{11}{\rm B}} + p + \nu (\bar{\nu})$ and $\nu  + {^{12}{\rm C}}    \rightarrow   {^{11}{\rm C}} + e^- + p$, and (5) the 15.11 MeV de-excitation line $\nu (\bar{\nu})+ {^{12}{\rm C}} \rightarrow \nu (\bar{\nu})+ {^{12}{\rm C}^\ast}$. Events from IBD and NC channels dominate over those from these channels (for a reference, see Table II in \cite{Lujan-Peschard:2014lta} and Table I in \cite{Lu:2016ipr}). Therefore, we neglect their contributions and focus on IBD and NC interactions.

In our previous work, the ratio used to probe neutrino mass hierarchy is the ratio of the total interactions of NC to those of IBD, which require reconstruction of the entire spectrum of SN neutrinos from detected events, $dN/dT^{\prime}$ and $dN/dE_{e^+}$. By exploring SN neutrino physics with a fixed set of parameters, we have shown that the capability of using SN neutrinos to probe neutrino parameters. In this work we would like to expand the parameter space of SN neutrinos by releasing the energy-equipartition condition and taking into account more combinations of mean energies of different flavors. we would also construct more realistic observables directly and explicitly related with detected events, $dN/dT^{\prime}$ and $dN/dE_{e^+}$, instead of the ratio of interactions. 
\label{NCreconstruct}

\section{Resolving Neutrino Mass Hierarchy}

Energy equipartition between all flavors is commonly assumed during the entire SN neutrino burst. However,  neutrino emissions actually evolve with time as the SN explodes. An important feature of the evolution of SN neutrino emissions is that the hierarchy of luminosities in accretion and cooling phases are reversed. For our present understanding, ${\mathcal L_{\nu_e}}\approx{\mathcal L_{\bar{\nu}_e}}>{\mathcal L_{\nu_x}}$ during the accretion phase and ${\mathcal L_{\nu_e}}\approx{\mathcal L_{\bar{\nu}_e}}<{\mathcal L_{\nu_x}}$ during the cooling phase. Therefore, while ${\mathcal L_{\nu_e}}\approx{\mathcal L_{\bar{\nu}_e}}$,  ${\mathcal L_{\nu_x}}/{\mathcal L_{\nu_e}}$ grows as SN neutrino emissions evolve from the accretion phase into the cooling phase. In addition to scan over plausible ranges for luminosity ratios, we check three specific scenarios corresponding to the accretion phase, the equipartition model, and the cooling phase, respectively, as in Table \ref{lumratio}. 

\begin{figure}[htbp]
	\begin{center}
	\includegraphics[width=0.3\textwidth]{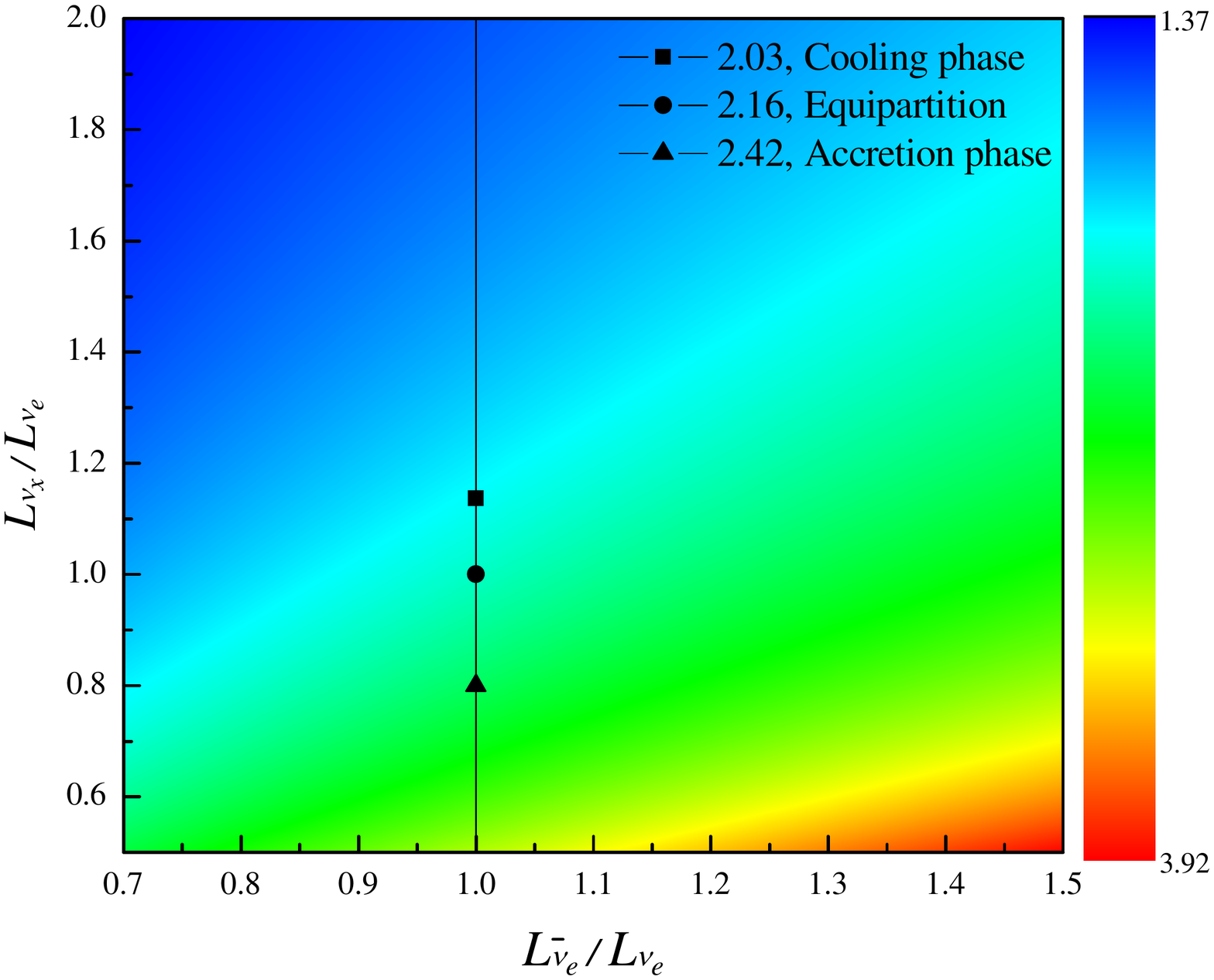}
	\includegraphics[width=0.3\textwidth]{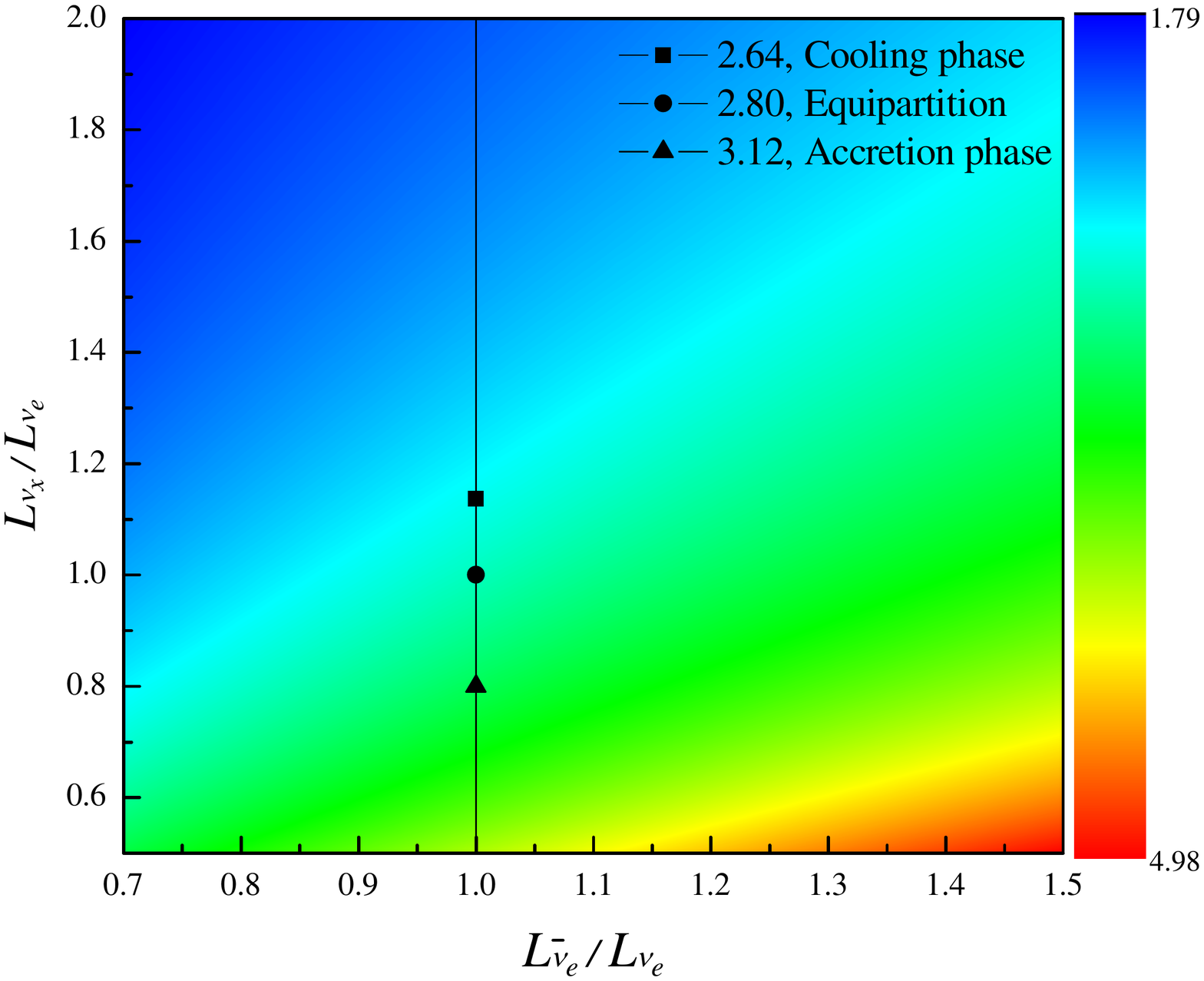}
	\includegraphics[width=0.3\textwidth]{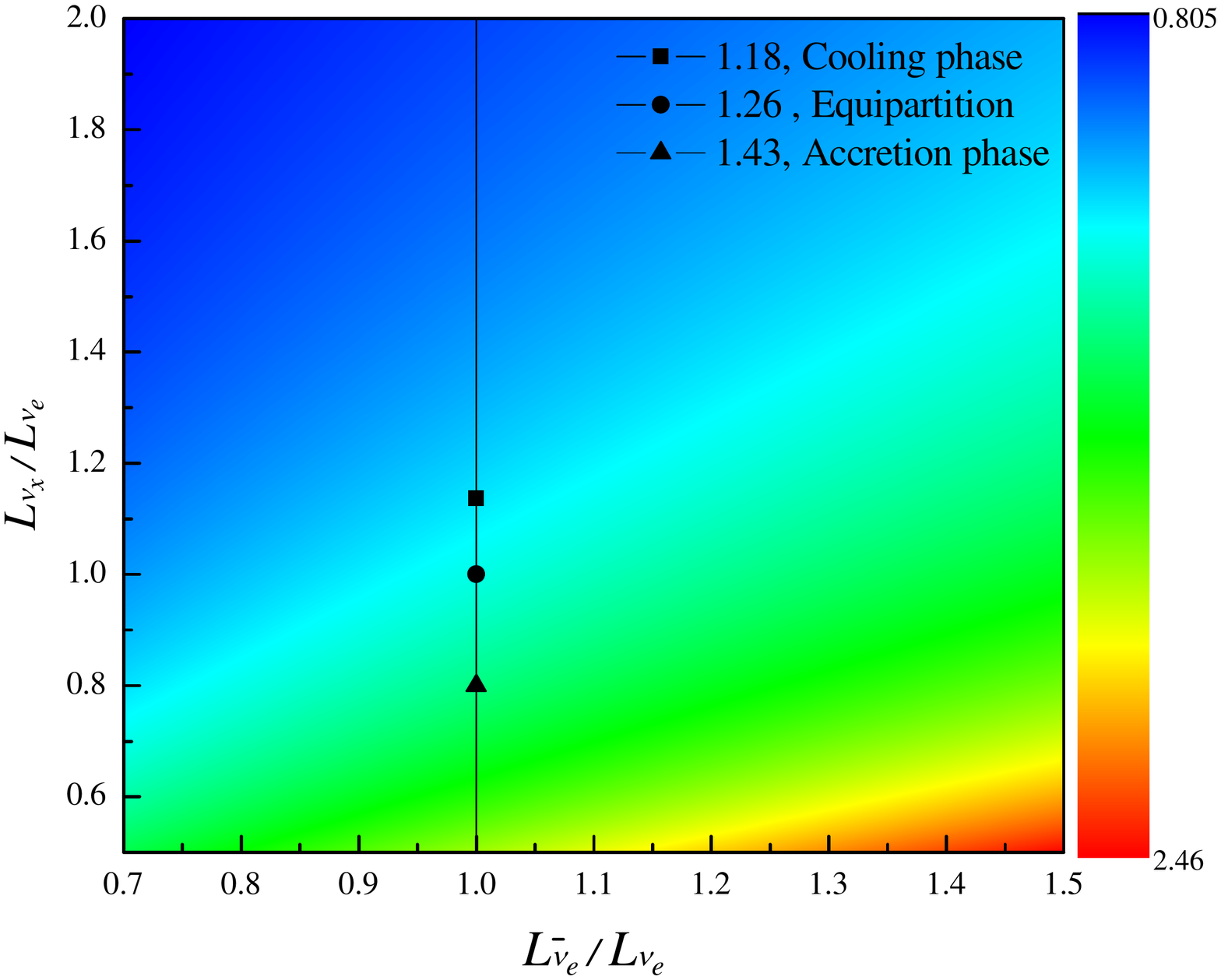}
	\includegraphics[width=0.3\textwidth]{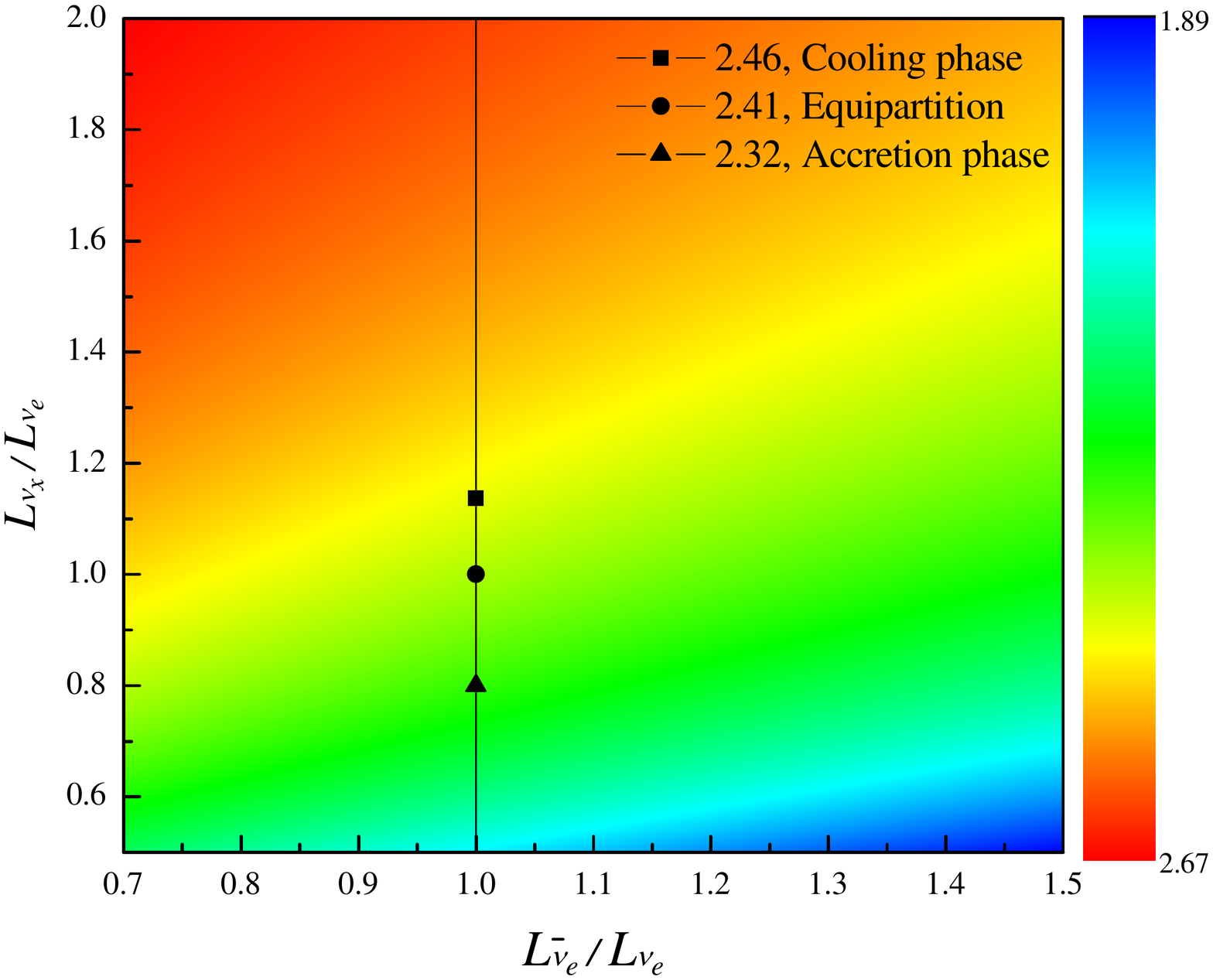}
	\includegraphics[width=0.3\textwidth]{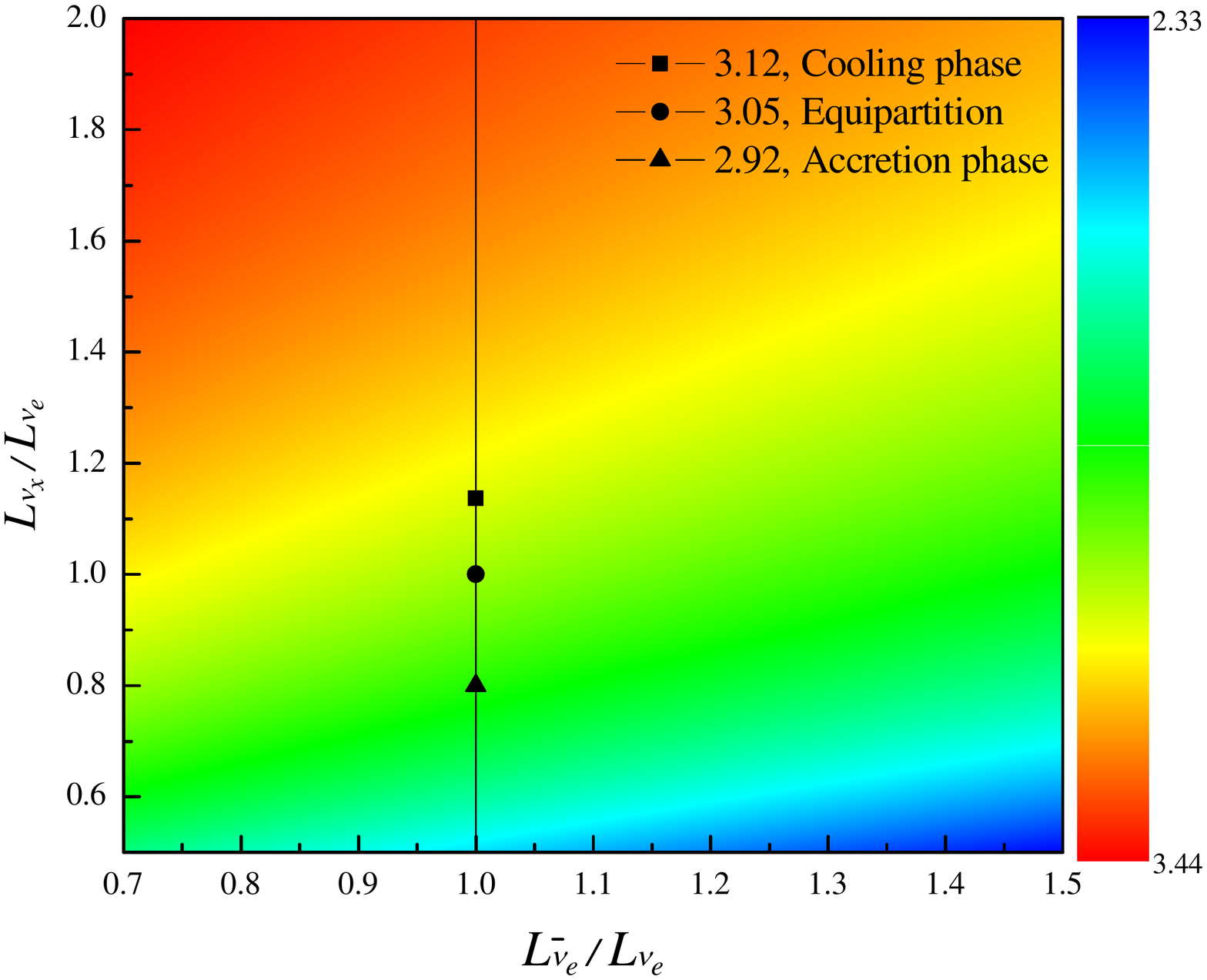}
	\includegraphics[width=0.3\textwidth]{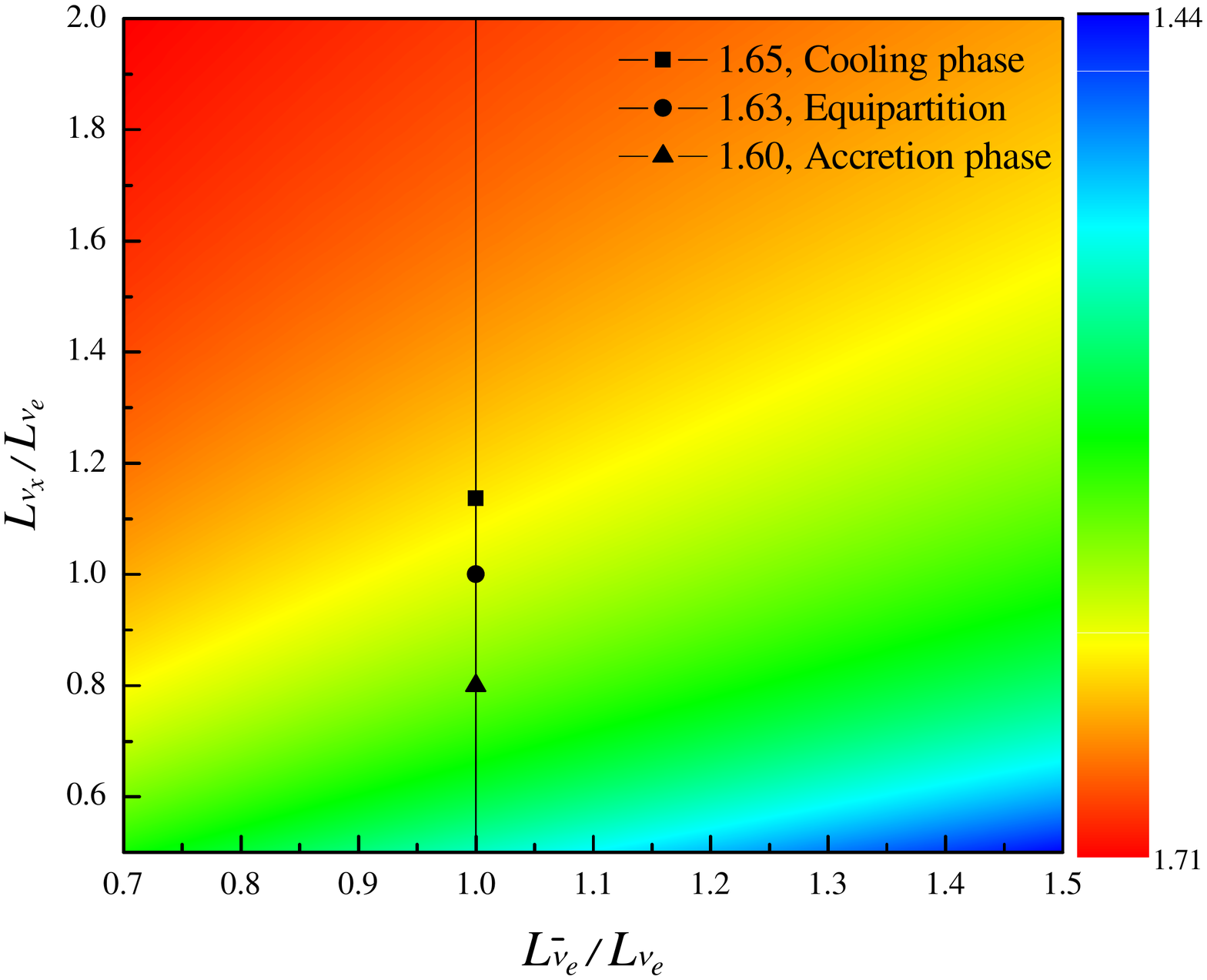}
	\caption{Ratios $R$ of IBD events to NC events in JUNO for different energy compositions in $\nu_e$, $\bar{\nu}_e$, and $\nu_x$ for the normal mass hierarchy on the upper panel and for the inverted mass hierarchy on the lower panel. From left to right, the three columns correspond to three different sets of neutrino mean energies of $(<E_{\nu_e}>, <E_{\bar{\nu}_e}>, <E_{\nu_x}>)$ taken from Sec. \ref{fluence}. $R$ values are scanned over $0.7<{\mathcal L_{\bar{\nu}_e}}/{\mathcal L_{\nu_e}}<1.5$ and $0.5<{\mathcal L_{\nu_x}}/{\mathcal L_{\nu_e}}<2.0$.}
	\label{fig:RJUNO}
	\end{center}
\end{figure}

The number of neutrino events inside the scintillator depends on $<E_{\alpha}>$'s and $\mathcal{L}_\alpha$'s, the SN neutrino parameters. In stead of assuming a standard SN neutrino emission model as in \cite{Lai:2016yvu}, we investigate the ratio of IBD events to NC events for different SN neutrino emission models by varying the SN neutrino parameters, $<E_{\alpha}>$'s and $\mathcal{L}_\alpha$'s, in calculating neutrino events inside the scintillator. We define $R$ to be the ratio of the total IBD events to the total NC events,
\begin{equation}
R=\frac{N_{\rm IBD}}{N_{\rm NC}}.
\end{equation}
For a specific set of $<E_{\alpha}>$'s, the event number of IBD is proportional to $\mathcal{L}_{\bar{\nu}_e}$ while that of NC is proportional to ${\mathcal L_{\nu_e}}+{\mathcal L_{\bar{\nu}_e}}+4{\mathcal L_{\nu_x}}$, the total luminosity. The ratio $R$ is then determined by the ratios of luminosities.

\begin{figure}[htbp]
	\begin{center}
	\includegraphics[width=0.3\textwidth]{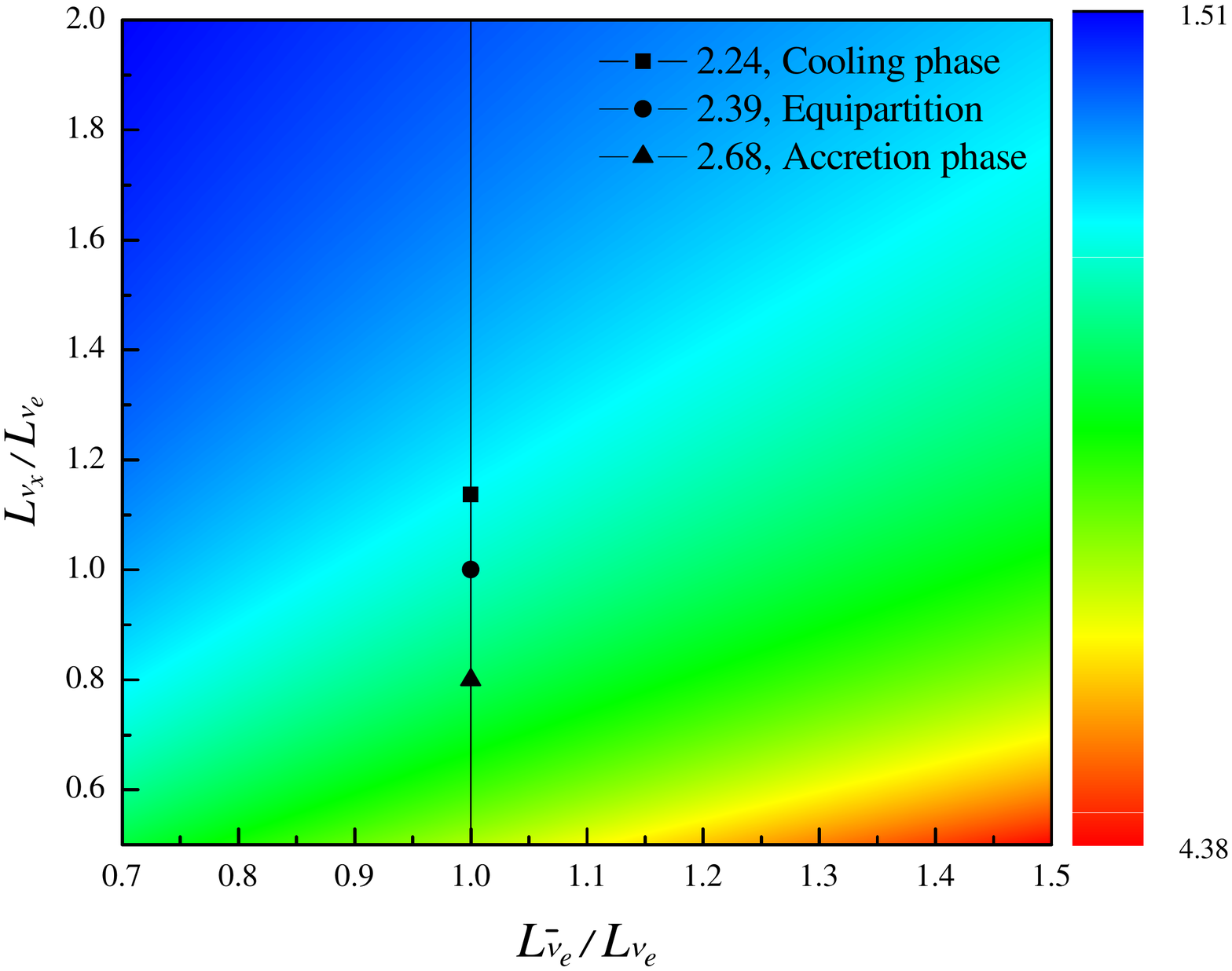}
	\includegraphics[width=0.3\textwidth]{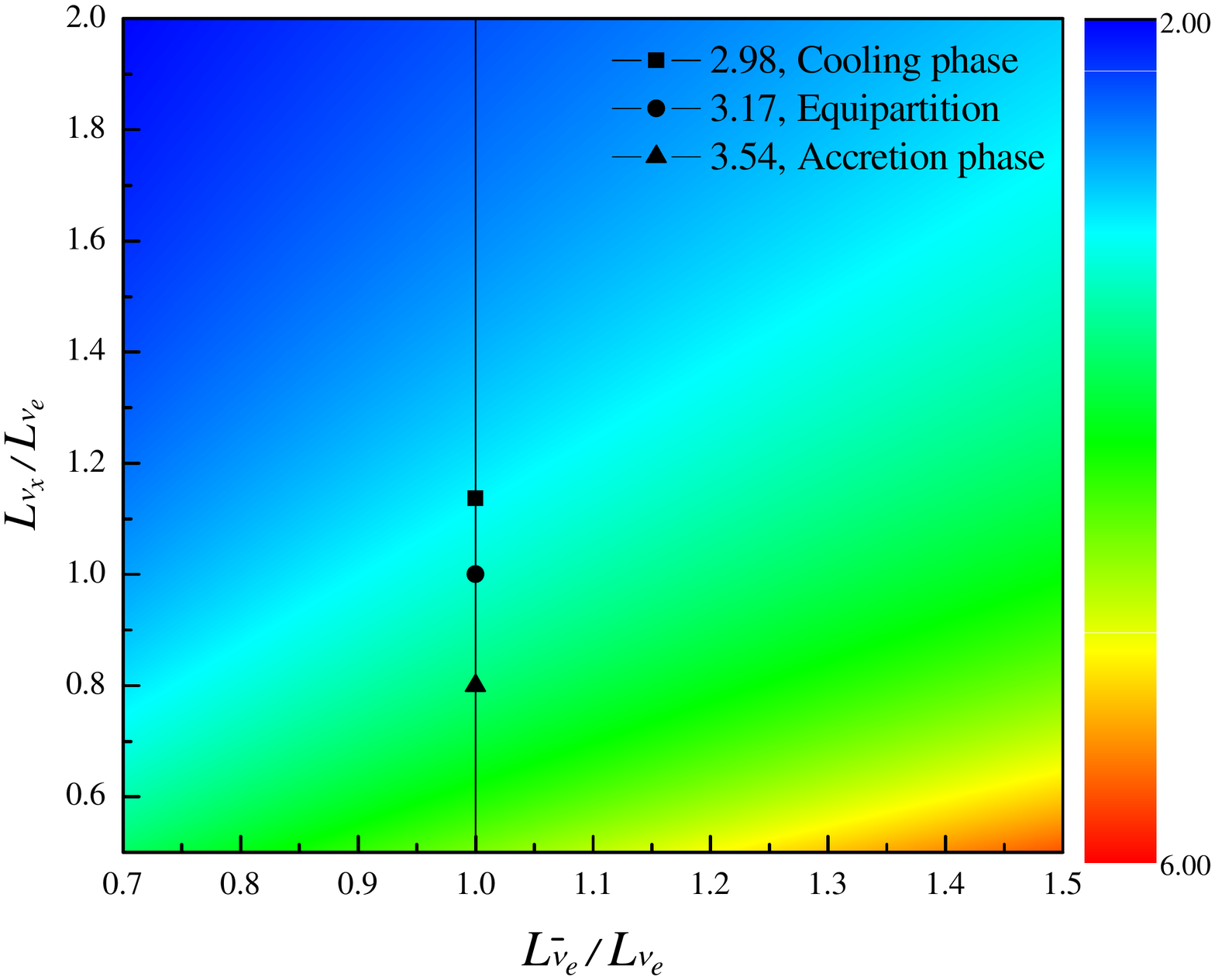}
	\includegraphics[width=0.3\textwidth]{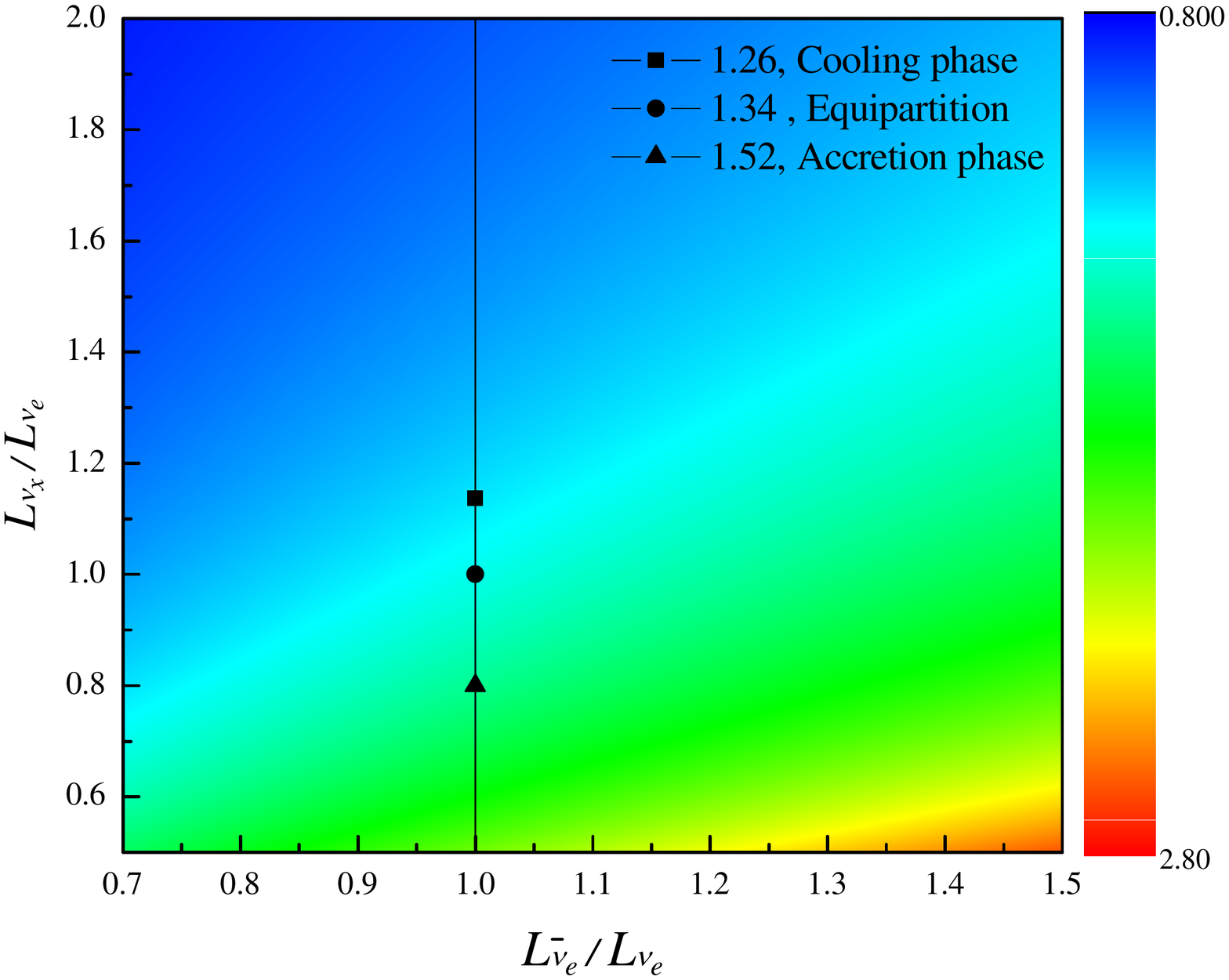}
	\includegraphics[width=0.3\textwidth]{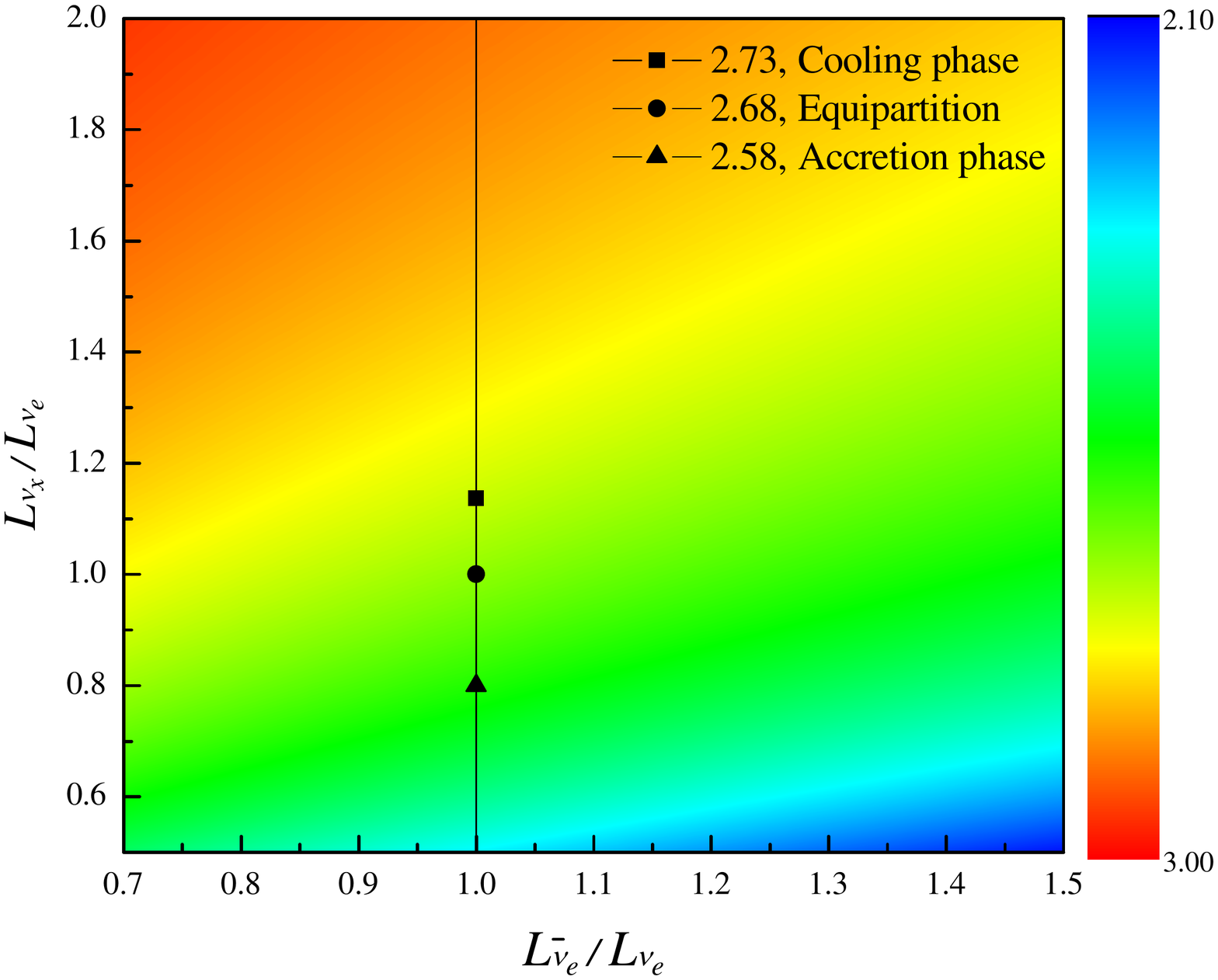}
	\includegraphics[width=0.3\textwidth]{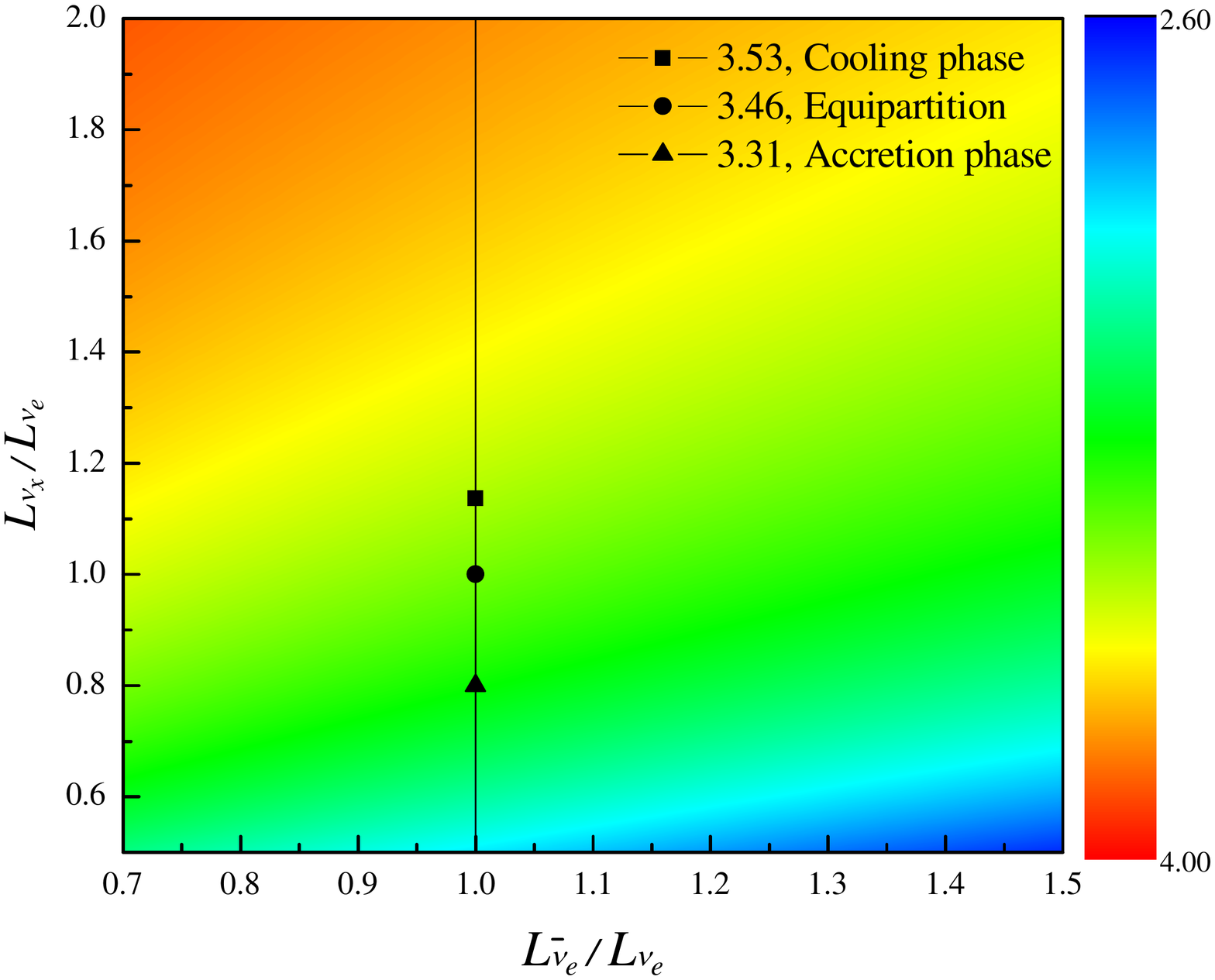}
	\includegraphics[width=0.3\textwidth]{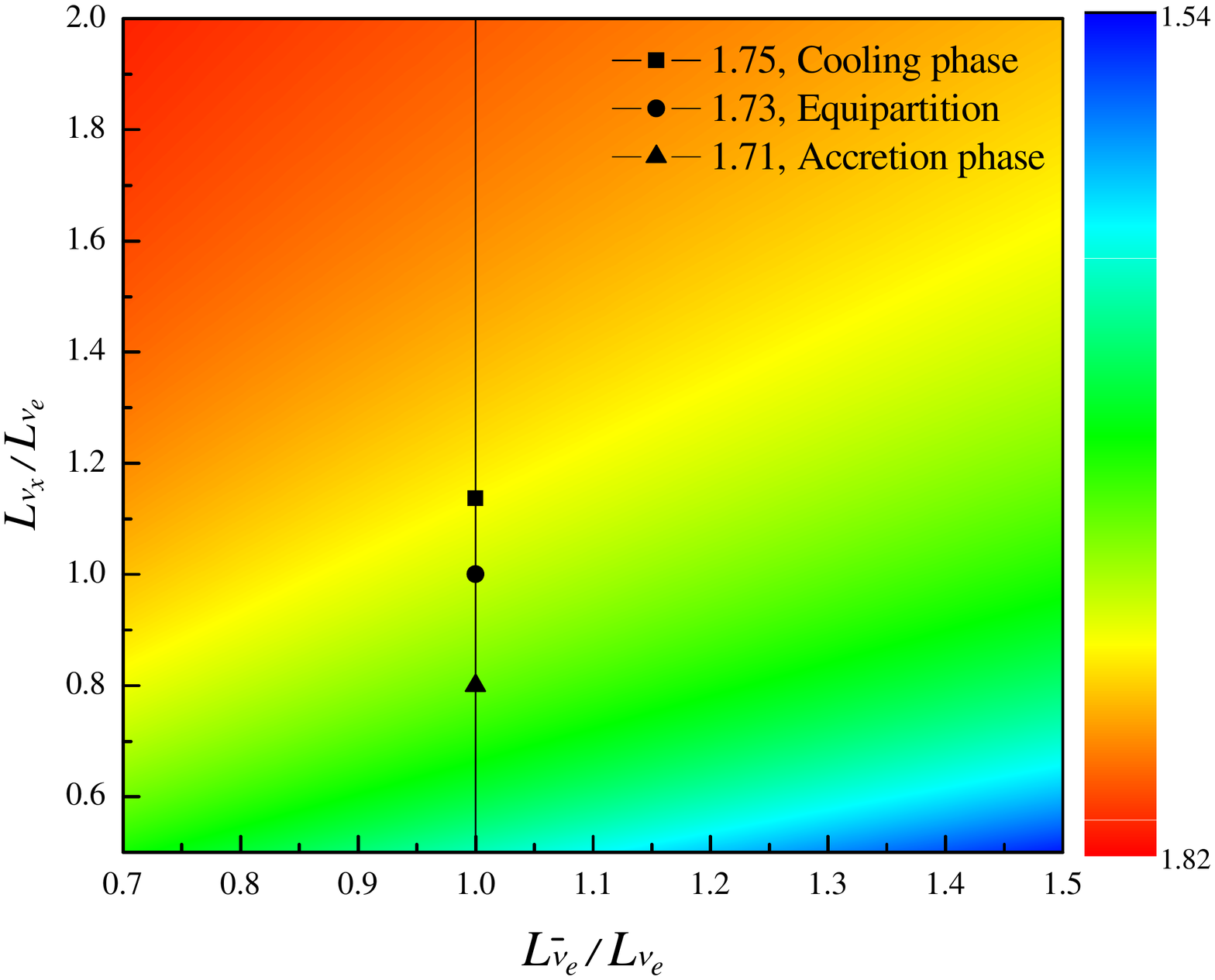}
	\caption{Ratios $R$ of IBD events to NC events in LENA for different energy compositions in $\nu_e$, $\bar{\nu}_e$, and $\nu_x$ for the normal mass hierarchy on the upper panel and for the inverted mass hierarchy on the lower panel. From left to right, the three columns correspond to three different sets of neutrino mean energies of $(<E_{\nu_e}>, <E_{\bar{\nu}_e}>, <E_{\nu_x}>)$ taken from Sec. \ref{fluence}. $R$ values are scanned over $0.7<{\mathcal L_{\bar{\nu}_e}}/{\mathcal L_{\nu_e}}<1.5$ and $0.5<{\mathcal L_{\nu_x}}/{\mathcal L_{\nu_e}}<2.0$.}
	\label{fig:RLENA}
	\end{center}
\end{figure}

While the $\nu p$ elastic scattering cross section is identical for all flavors and accounts for the total neutrino fluence, the NC spectrum should be the same for both neutrino mass hierarchies. Meanwhile, the IBD spectrum shall be different for neutrino flavor conversions inside the supernova are sensitive to neutrino mass hierarchy. As ${\mathcal L_{\nu_x}}/{\mathcal L_{\nu_e}}$ grows from smaller than one in the accretion phase to larger than one in the cooling phase, the fraction of $\nu_x$ flux to the total neutrino flux increases. Due to the dense matter inside the SN, the $\bar{\nu}_e$ flux is fully swapped with the $\bar{\nu}_x$ flux in the inverted hierarchy by the MSW effect. As a result, Eq. \ref{ebarIH} implies that $R$, the fraction of IBD events to NC events shall increase as the SN neutrino emission evolves from the accretion phase to the cooling phase for the inverted hierarchy. On the contrary, the fraction of $\bar{\nu}_e$ flux to the total neutrino flux decreases from the accretion phase to the cooling phase. Eq. \ref{ebarNH} then implies that $R$ shall decrease from the accretion phase to the cooling phase for the normal hierarchy.

\begin{table}[htbp]
\begin{center}
\begin{tabular}{lcc}\hline\hline
     & ${\mathcal L_{\bar{\nu}_e}}/{\mathcal L_{\nu_e}}$  & ${\mathcal L_{\nu_x}}/{\mathcal L_{\nu_e}}$   \\ \hline
Accretion Phase    & 1.00   &      0.80          \\ 
Energy Equipartition & 1.00  &     1.00          \\
Cooling Phase    & 1.00  &        1.14               \\ \hline
\end{tabular}
\end{center}
\caption{The luminosity ratios between different flavors for three specific scenarios.}
\label{lumratio}
\end{table}

For both normal and inverted hierarchies, we scan $R$ over the luminosity range of $0.5\leq{\mathcal L_{\nu_x}}/{\mathcal L_{\nu_e}}\leq2.0$ and $0.7<{\mathcal L_{\bar{\nu}_e}}/{\mathcal L_{\nu_e}}<1.5$ for three scenarios with mean energies presented in Sec. \ref{fluence}. The values of $R$ at JUNO \cite{An:2015jdp} and LENA \cite{Wurm:2011zn} detectors are shown in Fig. \ref{fig:RJUNO} and \ref{fig:RLENA}, respectively, in which the plots on the upper panel are for the normal hierarchy while those on the lower panel are for the inverted hierarchy. From left to right, the three columns correspond to three different sets of neutrino mean energies of $(<E_{\nu_e}>, <E_{\bar{\nu}_e}>, <E_{\nu_x}>)=(12~\rm MeV,~15~\rm MeV,~18~\rm MeV)$, $(12~\rm MeV,$ $~14~\rm MeV,~16~\rm MeV)$, and $(10~\rm MeV,~15~\rm MeV,~24~\rm MeV)$, respectively. 

Since ${\mathcal L_{\nu_e}}\approx{\mathcal L_{\bar{\nu}_e}}$, we can fix to ${\mathcal L_{\bar{\nu}_e}}/{\mathcal L_{\nu_e}}=1$ and explore how the $R$ value changes as ${\mathcal L_{\nu_x}}/{\mathcal L_{\nu_e}}$ grows. It is clearly seen, especially along the line at which ${\mathcal L_{\bar{\nu}_e}}/{\mathcal L_{\nu_e}}=1$, that the directions of the change of $R$ are opposed to each other for normal and inverted hierarchies. As the SN explosion evolves from the accretion phase to the cooling phase, ${\mathcal L_{\nu_x}}/{\mathcal L_{\nu_e}}$ increases from less than $1$ to greater than $1$ along the line from bottom up. Meanwhile, $R$ is decreasing for the normal hierarchy on the left panel and increasing for the inverted hierarchy on the right panel. Therefore, the neutrino mass hierarchy can be identified by measuring the change of $R$ from the accretion phase to the cooling phase. To illustrate this, we take a two-phase scenario to model the the time evolution of the SN neutrino emission as proposed in \cite{Lai:2016yvu}. The energy ratio between flavors in each phase are taken to be 
\begin{equation}
{\mathcal E}_{\nu_e,{\mathcal A}}:{\mathcal E}_{\bar{\nu}_e,{\mathcal A}}:{\mathcal E}_{\nu_x,{\mathcal A}}:{\mathcal E}_{\nu_e,{\mathcal  C}}:{\mathcal E}_{\bar{\nu}_e,{\mathcal  C}}:{\mathcal E}_{\nu_x,{\mathcal  C}} = 30:30:24:22:22:25,
\end{equation}
where ${\mathcal A}$ and ${\mathcal  C}$ denote the accretion and cooling phases and are marked by triangles and squares on the plots in Fig. \ref{fig:RJUNO} and \ref{fig:RLENA}, respectively. The corresponding values of ${\mathcal L_{\bar{\nu}_e}}/{\mathcal L_{\nu_e}}$ and ${\mathcal L_{\nu_x}}/{\mathcal L_{\nu_e}}$ are shown in Table \ref{lumratio}. Assuming a SN explosion with a total energy output of $\mathcal E=3\times10^{53}~\rm erg$ at a distance of $10~\rm kpc$, the values of $R$ and related event numbers are presented in Table \ref{junoevent} and \ref{lenaevent} for JUNO and LENA, rspectively. We note that, for the equipartition scenario marked by circles, the values in the table are calculated for the whole duration of the explosion with a total energy of $3\times10^{53}~\rm erg$. Hence the numbers are larger than those for the accretion and cooling phases. 

\begin{table}[htbp]
\begin{center}
\begin{tabular}{llrrrcccc}\hline\hline
               &   & \multicolumn{1}{c}{NC}  & \multicolumn{2}{c}{IBD} & \multicolumn{2}{c}{R} & \multicolumn{2}{c}{$\sigma_{\rm R}[10^{-2}]$} \\ \cline{4-9}
              &       &        &  IH    & NH   &   IH   &   NH  &   IH     & NH \\ \hline
                    & Accretion   &   1245    &  2888   &  3008       & 2.32  & 2.42   & 7.87  & 8.15     \\ 
(12, 15, 18)    & Equipartition     &  2493   & 6017  &  5383  & 2.41  & 2.16  & 5.75  & 5.23     \\
                  & Cooling  &  1223   & 3009  & 2480  & 2.46  & 2.03  & 8.35  & 7.09         \\  \hline
                  & Accretion     &  892  & 2600  & 2783  & 2.92  & 3.12  & 11.3  & 12.0      \\
(12, 14, 16)        & Equipartition        &  1775  & 5417  & 4973  & 3.05  & 2.80  & 8.35  & 7.75     \\ 
                  & Cooling  &  867  & 2708  & 2288  & 3.12  & 2.64  & 12.2  & 10.5     \\ \hline
                  & Accretion     &  2265  & 3633  & 3237  & 1.60  & 1.43  & 4.29  & 3.92      \\
(10, 15, 24)    & Equipartition     &  4637  & 7569  & 5861  & 1.63  & 1.26  & 3.04  & 2.48        \\ 
                  & Cooling  &  2299  & 3785  & 2719  & 1.65  & 1.18  & 4.35  & 3.35     \\ \hline

\end{tabular}
\end{center}
\caption{Numbers of IBD and NC events in three specific scenarios listed in Table \ref{lumratio} for different sets of the average mean energies $(<E_{\nu_e}>, ~<E_{\bar{\nu}_e}>, ~<E_{\nu_x}>)$ at JUNO}
\label{junoevent}
\end{table}

\begin{table}[htbp]
\begin{center}
\begin{tabular}{llrrrcccc}\hline\hline
               &   & \multicolumn{1}{c}{NC}  & \multicolumn{2}{c}{IBD} & \multicolumn{2}{c}{R} & \multicolumn{2}{c}{$\sigma_{\rm R}[10^{-2}]$} \\ \cline{4-9}
                       &                          &               &    IH        &      NH      &      IH       &   NH      &      IH     &     NH      \\ \hline
                       & Accretion          &   2522    &    6497    &    6766      &    2.58     &   2.68    &   6.05    &   6.26     \\ 
(12, 15, 18)    & Equipartition     &   5059     &   13536   &   12110    &     2.68     &   2.39    &   4.41    &   4.01     \\
                      & Cooling              &   2483    &    6768    &    5578     &     2.73     &   2.25    &   6.39    &   5.42         \\  \hline
                      & Accretion           &   1768     &   5849    &    6261     &     3.31     &   3.54    &   8.98     &   9.54      \\
(12, 14, 16)    & Equipartition      &   3523    &   12185  &    11186    &     3.46     &   3.17    &   6.62     &   6.13     \\ 
                      & Cooling              &   1724    &    6093   &     5147     &     3.53     &   2.99    &   9.64     &   8.31     \\ \hline
                      & Accretion           &    4785   &    8173    &    7282     &     1.71     &   1.52    &   3.11      &   2.83      \\
(10, 15, 24)    & Equipartition     &    9809    &   17027  &    13185    &     1.74     &   1.34    &   2.20     &   1.79        \\ 
                      & Cooling              &   4866    &    8514   &    6116      &     1.75     &   1.26    &   3.14      &  2.41     \\ \hline

\end{tabular}
\end{center}
\caption{Numbers of IBD and NC events in three specific scenarios listed in Table \ref{lumratio} for different sets of the average mean energies $(<E_{\nu_e}>, ~<E_{\bar{\nu}_e}>, ~<E_{\nu_x}>)$ at LENA} 
\label{lenaevent}
\end{table}

When the mean energy distribution becomes more hierarchical, the event numbers of both NC and IBD become larger because more neutrinos are shifted to the high energy tail and the the cross sections are larger at higher energies. Since IBD event number depends on the the flux of Eq. (\ref{ebarNH}) or (\ref{ebarIH}) while NC event number depends on the the total flux of $F_e+F_{\bar{e}}+4F_x$, the NC event number grows more than the IBD one. Therefore, we find that $R$ is smaller when the mean energy distribution is more hierarchical in spite of the mass hierarchy as shown in Table \ref{junoevent} and \ref{lenaevent}. For the inverted hierarchy, the IBD event number is determined by $F^0_x$ from Eq. (\ref{ebarIH}). Since the energy fraction of the $\nu_x$ in the accretion phase is smaller than that in the cooling phase, ${\mathcal E}_{\nu_x,{\mathcal A}}/({\mathcal E}_{\nu_e,{\mathcal A}}+{\mathcal E}_{\bar{\nu}_e,{\mathcal A}}+4{\mathcal E}_{\nu_x,{\mathcal A}})<{\mathcal E}_{\nu_x,{\mathcal  C}}/({\mathcal E}_{\nu_e,{\mathcal  C}}+{\mathcal E}_{\bar{\nu}_e,{\mathcal  C}}+4{\mathcal E}_{\nu_x,{\mathcal  C}})$, the value of $R$ in the accretion phase should be smaller than that in the cooling phase, $R_{\mathcal A}<R_{\mathcal C}$, for the inverted hierarchy. On the contrary, the energy fraction of the $\bar{\nu}_e$ in the accretion phase is larger than that in the cooling phase, ${\mathcal E}_{\bar{\nu}_e,{\mathcal A}}/({\mathcal E}_{\nu_e,{\mathcal A}}+{\mathcal E}_{\bar{\nu}_e,{\mathcal A}}+4{\mathcal E}_{\nu_x,{\mathcal A}})~>~{\mathcal E}_{\bar{\nu}_e,{\mathcal  C}}/({\mathcal E}_{\nu_e,{\mathcal  C}}+{\mathcal E}_{\bar{\nu}_e,{\mathcal  C}}+4{\mathcal E}_{\nu_x,{\mathcal  C}})$. Meanwhile, one has $|{\mathcal E}_{\nu_x,{\mathcal A}}/({\mathcal E}_{\nu_e,{\mathcal A}}+{\mathcal E}_{\bar{\nu}_e,{\mathcal A}}+4{\mathcal E}_{\nu_x,{\mathcal A}})-{\mathcal E}_{\nu_x,{\mathcal  C}}/({\mathcal E}_{\nu_e,{\mathcal  C}}+{\mathcal E}_{\bar{\nu}_e,{\mathcal  C}}+4{\mathcal E}_{\nu_x,{\mathcal  C}})|<|{\mathcal E}_{\bar{\nu}_e,{\mathcal A}}/({\mathcal E}_{\nu_e,{\mathcal A}}+{\mathcal E}_{\bar{\nu}_e,{\mathcal A}}+4{\mathcal E}_{\nu_x,{\mathcal A}})-{\mathcal E}_{\bar{\nu}_e,{\mathcal  C}}/({\mathcal E}_{\nu_e,{\mathcal  C}}+{\mathcal E}_{\bar{\nu}_e,{\mathcal  C}}+4{\mathcal E}_{\nu_x,{\mathcal  C}})|$. As a result, $R_{\mathcal A}>R_{\mathcal C}$ is inferred from Eq. \ref{ebarNH} for the normal hierarchy. 

\begin{figure}[htbp]
	\begin{center}
	\includegraphics[width=0.3\textwidth]{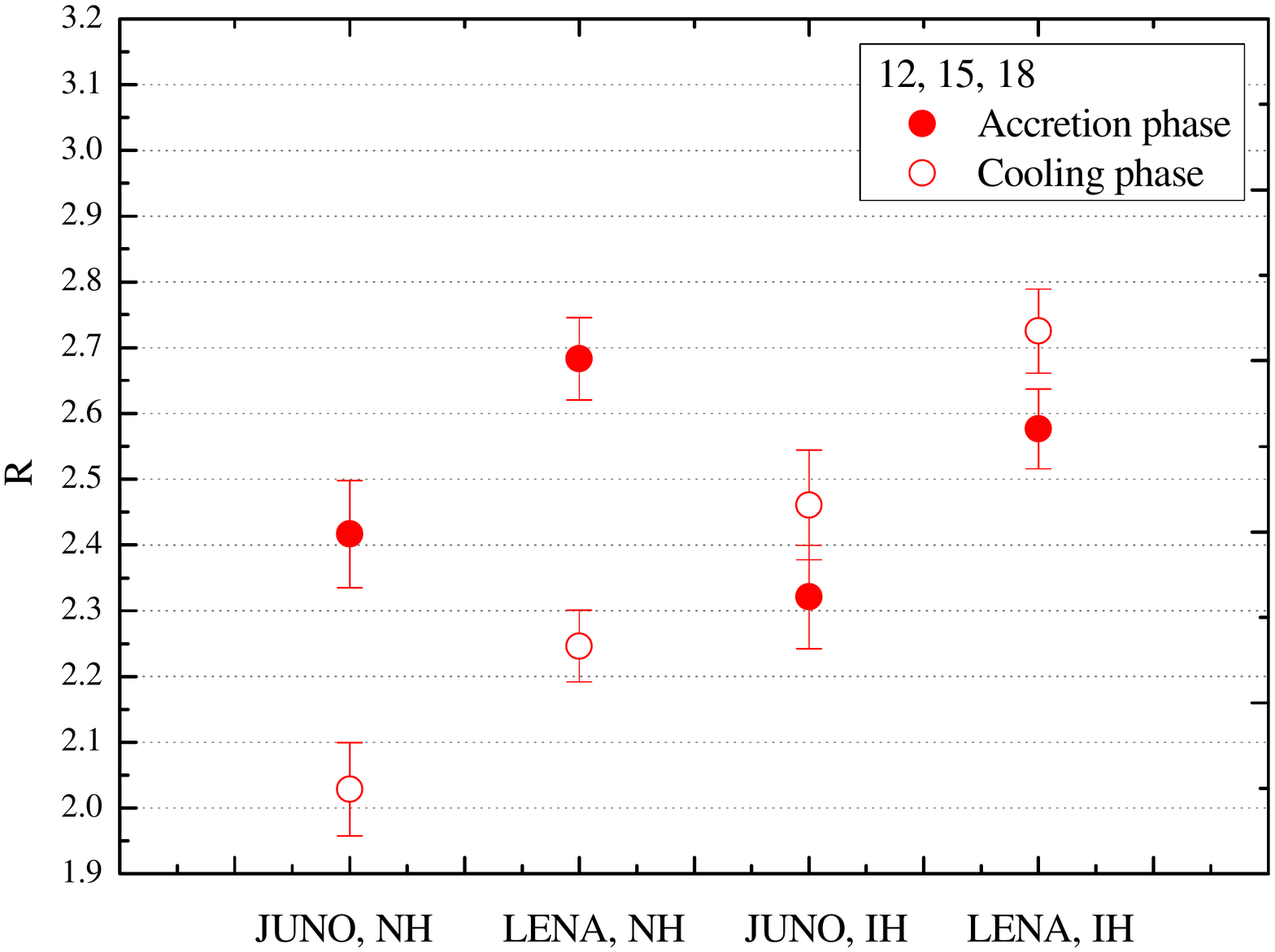}
	\includegraphics[width=0.3\textwidth]{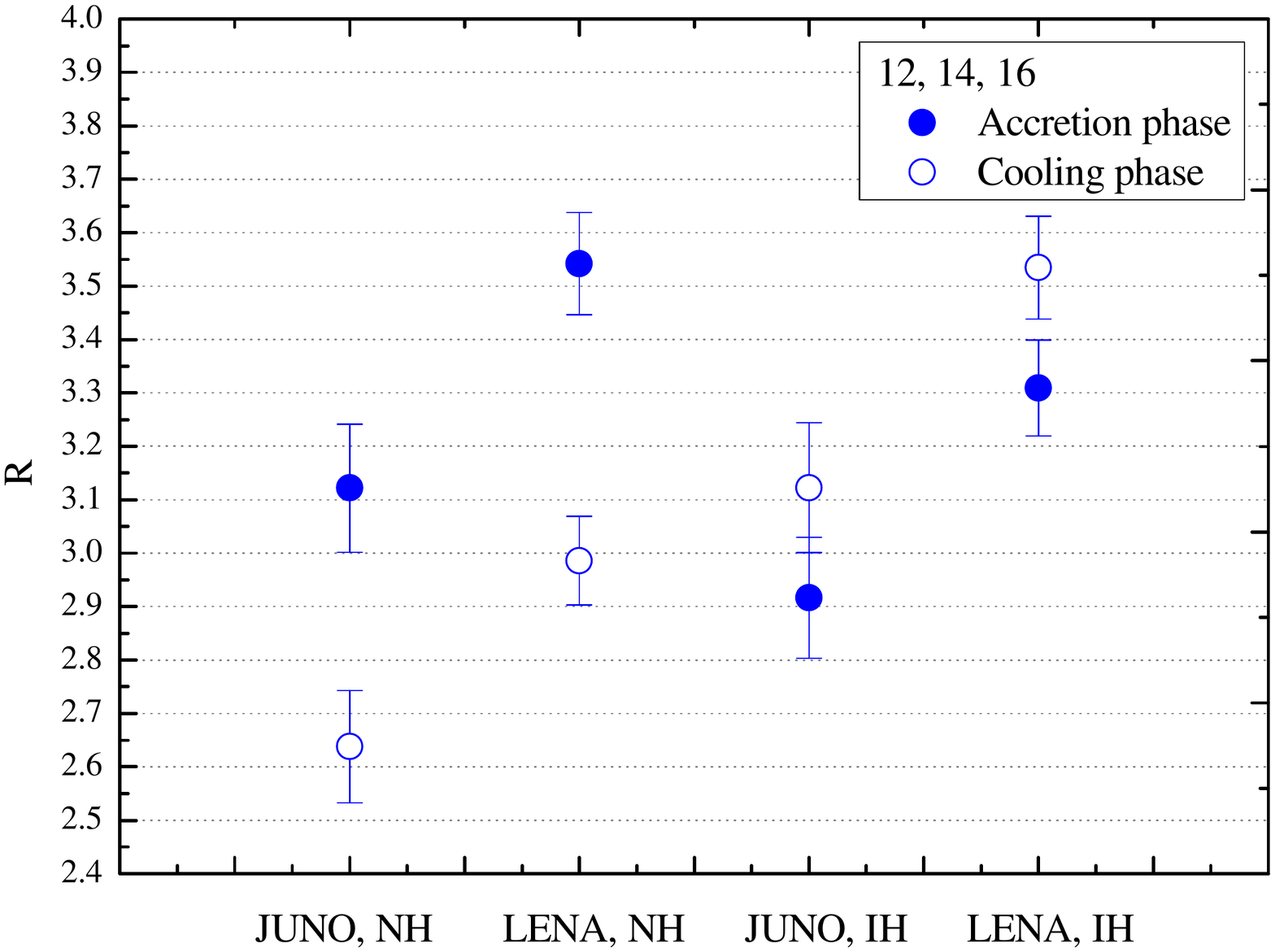}
	\includegraphics[width=0.3\textwidth]{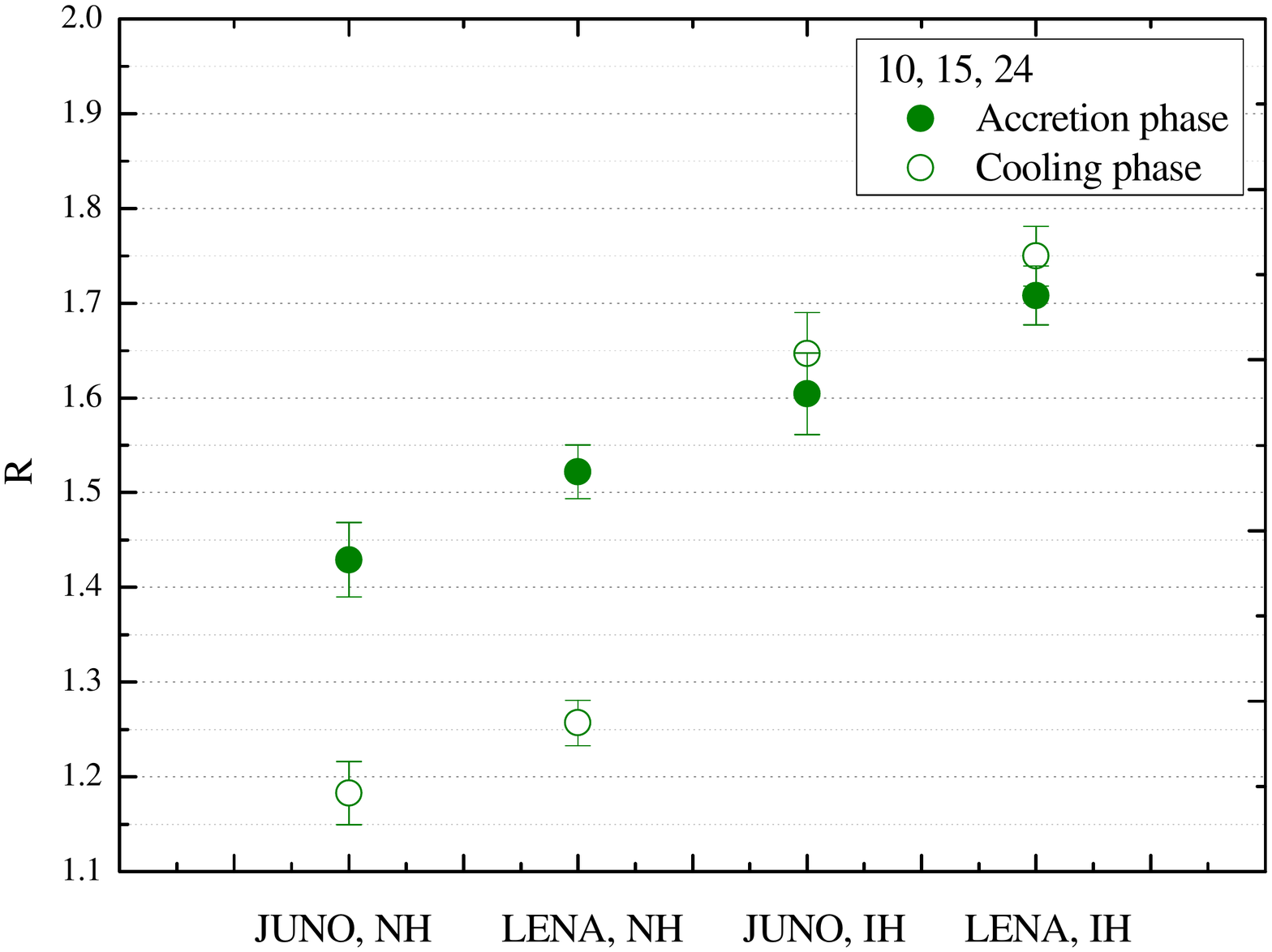}
	\caption{Expected $R$ values and uncertainties at JUNO and LENA detectors for both mass hierarchies. Each plot corresponds to a set of mean energies of different flavors denoted at the upper right corner in MeV. }
	\label{fig:sigmaR}
	\end{center}
\end{figure}

To determine whether $R_{\mathcal A}>R_{\mathcal C}$ or $R_{\mathcal A}<R_{\mathcal C}$ requires clear discrimination between $R_{\mathcal A}$ and $R_{\mathcal C}$. This can be achieved by requiring the ranges of $R_{\mathcal A}$ and $R_{\mathcal C}$ do not overlap. The ranges of $R$ at JUNO and LENA for the three sets of mean energies are shown in Fig. \ref{fig:sigmaR}. One can see that, with the measurements of JUNO, NMH can be identified if it is normal but it cannot be determined if it is inverted, since the ranges of $R_{\mathcal A}$ and $R_{\mathcal C}$ at JUNO overlap for all the three sets of mean energies in the inverted hierarchy. At LENA, $R_{\mathcal A}$ and $R_{\mathcal C}$ are clearly discriminated in between, except for the mean energies of $(10~\rm MeV,~15~\rm MeV,~24~\rm MeV)$ in the inverted hierarchy. Therefore, LENA is capable of determining the NMH for less hierarchical mean energy distributions. Moreover, one can also infer that, even in the most hierarchical distribution of mean energies, the ranges of $R_{\mathcal A}$ and $R_{\mathcal C}$ can finally be separated as long as the detector mass is large enough to collect enough events. From Tables \ref{junoevent} and \ref{lenaevent}, the difference between $R_{\mathcal A}$ and $R_{\mathcal C}$ is smaller in the inverted hierarchy than in the normal hierarchy. This is the reason why the normal hierarchy is easier to be identified than the inverted hierarchy.

We note that $R$ values for the same model parameters are different at JUNO and LENA. As  shown in Table \ref{Detector},  the scintillation materials in the two detectors are different resulting in different Birk's constants. When applying the same cut of $0.2 \rm MeV$ to the quenched signal $T^{\prime}$, the corresponding proton recoil $T_{\rm min}$'s are different for the two detectors. The higher $T_{\rm min}$ for LENA indicates that fewer fraction of proton recoils are picked as NC events such that the $R$ values are larger for LENA than for JUNO while applied to the same mean energies and luminosity ratios. 

\begin{table}[htbp]
\begin{center}
\begin{tabular}{lccccccc}\hline\hline
                  & Mass  &            $N_{\rm p}$ & $k_B$      &  $T^{\prime}_{\rm min}$    &   $T_{\rm min}$ \\ 
                  & [kton]   &           $[10^{31}]$  & [cm/MeV] &  [MeV]  & [MeV]   \\ \hline
JUNO        &  20 &           144      &     0.00759  &       0.2      &   0.93       \\
LENA         &  44 &            325     &     0.010      &        0.2      &  1.02            \\ \hline

\end{tabular}
\end{center}
\caption{Detector properties for the future scintillation detectors considered in this work. Masses of scintillation materials, corresponding Birks constants ($k_{\rm B}$), numbers of free protons ($N_{\rm p}$), thresholds of quenched energy ($T^{\prime}_{\rm min}$), and thresholds of proton recoil energy ($T_{\rm min}$).}
\label{Detector}
\end{table}

\section{Summary and Conclusions}

We have described how to identify the neutrino mass hierarchy by measuring the time variation of the SN neutrino events inside scintillation detectors. IBD events and NC events at scintillation detectors are taken to define the event ratio $R$, which can be calculated for given mean energies of and luminosity ratios between different flavors with detector parameters. Our knowledge of the time evolution of SN neutrino emissions indicates that, as the SN explosion evolves from the accretion phase to the cooling phase, the ratio $R$ of IBD events to NC events shall decrease for the normal mass hierarchy and increase for the inverted mass hierarchy. We not only calculate $R$ over physically plausible ranges of luminosity ratios between flavors but also evaluate statistical uncertainties of $R$ arising from measurements for an illustrative model of the SN neutrino emissions. This clarify the detector capability for resolving the neutrino mass hierarchy. 

We have performed our analysis with three different sets of mean energies of flavors in two scintillation detectors, JUNO and LENA, and found that, as ${\mathcal L_{\nu_x}}/{\mathcal L_{\nu_e}}$ grow, $R$ is increasing for the inverted hierarchy and decreasing for the normal hierarchy in spite of the mean energies of different flavors. With a two-phase scenario to model the SN neutrino emission, we have presented the IBD and NC events, $R$'s, and, $\sigma_R$'s, deviations of $R$, numerically to illustrated that $R$ does change in opposite directions as SN neutrinos evolves from the accretion phase to the cooling phase. The detector capability for different mass hierarchies has also been checked and compared for the two detectors. and we have found that the neutrino mass hierarchy will be easier to be identified if it is normal.

A SN neutrino model can be characterized by the luminosity, ${\mathcal L}$, the mean energy, $<E>$, and the shape parameter, $\eta$ for each flavor. Simulations (see \cite{Mirizzi:2015eza} for a review) have shown that $\eta$'s do not vary much between $2$ and $3$. Therefore, a common $\eta$ is assumed throughout this work for all three flavors during both phases for simplicity. In principle, all the parameters, ${\mathcal L}$, $<E>$, and $\eta$, vary with time. To incorporate the time-dependence of these SN parameters, one can use simulation data of SN neutrino emissions to obtain the event rate, $dN/dt$. Besides of IBD and NC interactions, SN $\nu_e$ flux can also be measured in the liquid scintillator \cite{Laha:2014yua}. In future studies, we shall work on event rates for different channels of neutrino interactions in various detector to explore the SN explosion with all three flavors of neutrinos taken into account.

\section*{Acknowledgements}
We thank  G.-L. Lin and Jason Leung for helpful discussions and comments. This work is partly supported by the Ministry of Science and Technology, Taiwan, under Grants No. MOST 106-2112-M-182-001.

\end{document}